\documentclass[epjST]{svjour}
\usepackage{graphicx}
\begin{document}
\title{Fractal transit networks: self-avoiding walks and L\'evy  flights}
 %\subtitle{}
\author{Christian von Ferber\inst{1}\fnmsep\thanks{\email{C.vonFerber@coventry.ac.uk}} \and
Yurij Holovatch\inst{3}\fnmsep\thanks{\email{hol@icmp.lviv.ua}}}
\institute{Applied Mathematics Research Centre, Coventry University,
Coventry CV1 5FB, UK \&
Heinrich-Heine Universit\"at D\"usseldorf, D-40225 D\"usseldorf, Germany
\and
 Institute for Condensed Matter Physics, National Acad. Sci. of Ukraine,
 79011 Lviv, Ukraine}
\abstract{
Using data on the Berlin public transport network, the pre\-sent study extends
previous observations of fractality within
public transport routes by showing that also the distribution of
inter-station distances along routes displays non-trivial power law behaviour.
This indicates that the routes may in part
also be described as L\'evy-flights. The latter property may result
from the fact that the routes are planned to adapt to fluctuating
demand densities throughout the served area. We also relate this to optimization
properties of L\'evy flights.
} %end of abstract
\maketitle
\section{Introduction}
\label{I}

Transit networks are fascinating objects to study. While each network and every element in the
network needs to be adapted to a specific situation in the urban agglomeration, it turns out that
nonetheless general  construction principles prevail. The latter fact has been evidenced by
numerous studies using in particular methods of statistical physics, random graph theory, and
other approaches. One of the main fields contributing to this analysis is complex network
science \cite{small_world,complex} (see \cite{Berche12,Holovatch11} for a review). Applying the tools
of complex network theory it was shown, that transit networks
(in this paper we consider urban public transportation networks, PTN) share
common features of other natural and man-made complex networks. In particular,
they are strongly correlated compact structures
(so-called small worlds) \cite{small_world}, that often possess power law scale-free behaviour \cite{vonFerber}.
It has been shown, that they are resilient against random failures while being vulnerable to targeted
attacks \cite{Holovatch11,Berche09}. The study of fractal properties of PTN has a long history
\cite{Thibault87,Frankhauser90,Benguigui91,Benguigui92,Kim03}. The prevailing
part of these former studies concern the density of stations or the total length of track
as function of the distance from the center of the network. In this way PTNs and rail
networks of various cities and regions were analyzed.
Thibault and Marchand \cite{Thibault87}  studied the fractal
properties of different subnetworks within Lyon (regions I,II, III). The fractal dimensions
found were for the suburban rail I: 1.64, II: 1.66, III: 1.88; for the public bus service: 1.00, 1.09,
1.45; the authors also analyse the drainage utility with results: 1.21, 1.30, and 1.79.
Frankhauser \cite{Frankhauser90} obtained $1.58$ for
Stuttgart's rail system.
Benguigui and Daoud \cite{Benguigui91} obtained the fractal dimension for
the Paris railway transportation network as 1.47. In relation to this
result they also discuss DLA and star polymers as known models that
lead to fractal structure. Further results for fractal dimensions were  given by Benguigui \cite{Benguigui92}
for several Rhine towns and Moscow railways: $1.70\pm 0.05$,
Paris metro: $1.80\pm 0.05$.
The fractal dimension of Seul PTN has been estimated as  1.50 for stations
and 1.35 for tracks  \cite{Kim03}.
In a complementary approach, the present authors measured the mean distance
as a function of number of stations traveled \cite{vonFerber}.
For different Berlin subnetworks a fractal dimension in the range
of 1.04 -- 1.22 was found. The Paris metro data leads to a
fractal dimension 1.22 when excluding short distance
contributions.

Some of the results cited above appear to be compatible with the interpretation
of the routes as two dimensional self-avoiding walks (SAWs) with a fractal
dimension of $4/3$ \cite{Nienhuis82}. However, the results appear to vary considerably.
Here, in addition to the fractal dimension of the transit routes we intend to analyse
the distributions of inter-station distances of consecutive stations. We will show
that these distributions appear to have heavy (power-law) tails compatible with
a L\'evy-flight model for the transit routes. The latter approach further presents
a possible explanation for the deviations from the SAW behaviour observed.

A L\'evy-flight model is characterised by the distribution function
of the step lengths $\ell$ of the otherwise random successive steps.
For a specific L\'evy-flight the complementary (the 'tail') of the
cumulative distribution function $P(\ell)$ of the step lengths
$\ell$ follows a power law decay
\begin{equation} \label{1}
P(\ell)\sim \ell^{-\mu}.
\end{equation}
where the exponent $\mu$ characterises the distribution. Note that
that exponents greater than 2 belong to the Gaussian Central Limit
Theorem (CLT) domain of attraction due to the fact that for an
exponent $\mu> 2$ the probability density function (PDF) has finite
variance. Exponents $\mu<2$, however, belong to the L\'evy CLT
domain of attraction. The critical exponent $\mu=2$ is more delicate
and requires the analysis of the growth of the associated truncated
variance function.

In the following, we first describe the empirical data as they were drawn from the
Berlin PTN. Then, we analyze the distribution of the inter-station distances along the routes
and propose a description of PTN routes  and journeys in terms of self-avoiding L\'evy flights.
This is supported by the fact that the scaling of distance traveled as function of the number of
stations deviates from SAWs but is in line with L\'evy flight-like behaviour.

\section{Data analysis}
\label{II}

PTNs are often discussed without
reference to their geographical embedding. The fact that this
subject is left aside by most studies of
PTNs with respect to their complex network behaviour, is due mainly
to the lack of easily accessible data on the locations of stations
and routes. For the present work we have analyzed such data for
stations of the Berlin PTN. Our database consists of 2992 stations that belong to 211 routes.
Furthermore, we have taken into account different means of public
transit, that include subway (U-bahn), tram, bus, and surface high
speed transport (S-bahn). The data was collected in the frames of
a larger project \cite{Berche12,vonFerber}  including a comparative analysis of PTN vulnerability
of fourteen large cities  \cite{Holovatch11,Berche09}.
Here, we complement
available information about the Berlin PTN topology by the information
about the geographical location of each single station. To this end,
Cartesian coordinates for each station were extracted from an
on-line map provided by the Berlin public transport operator. We note that using this
procedure slight deviations between distances calculated on the spherical surface and
those derived from the projection to  a flat surface with Cartesian coordinates occur.
These will be neglected below. With the data on the topological
and geographical properties of the PTN at hand we are in the position to analyze how
these properties are related.

A short glance at the map of a public transport network immediately
reveals that transport lines do in general not follow the shortest
"straight ahead" route between their two end points. Thus the
question arises if there is any other underlying structure or
principle characterizing the observed behaviour of geographically
embedded transport routes. Let us analyse transport routes with
respect to their fractal properties in terms of random walks,
self-avoiding walks and L\'evy flights. In former studies
\cite{vonFerber} we have analyzed the dependence of the mean square
distance between stations $\langle R^2 \rangle$ as function of the
number $N$ of stations traveled. The following power law scaling was
found:
\begin{equation} \label{2}
\langle R^2 \rangle \sim N^{2\nu}
\end{equation}
with the exponent $\nu$ ranging from $\nu = 0.82$ for
bus routes to $\nu = 0.9$ and $0.96$ for subway and tram
routes. The S-bahn data is distorted due to a ring structure
within this sub-network. The obtained values of the exponent lie
in the region $3/4 \leq \nu \leq 1$.
These results are to be compared with the behaviour of self-avoiding walks (SAW)
in two dimensions characterised by an exponent $\nu=3/4 = 0.75$ which corresponds
to the exact result for the fractal dimension of $D_{SAW}=4/3$ \cite{Nienhuis82}.
This observation leads to the hypothesis that the PTN routes may follow  SAW statistics.

\begin{figure}
\centering
\includegraphics[width=57mm]{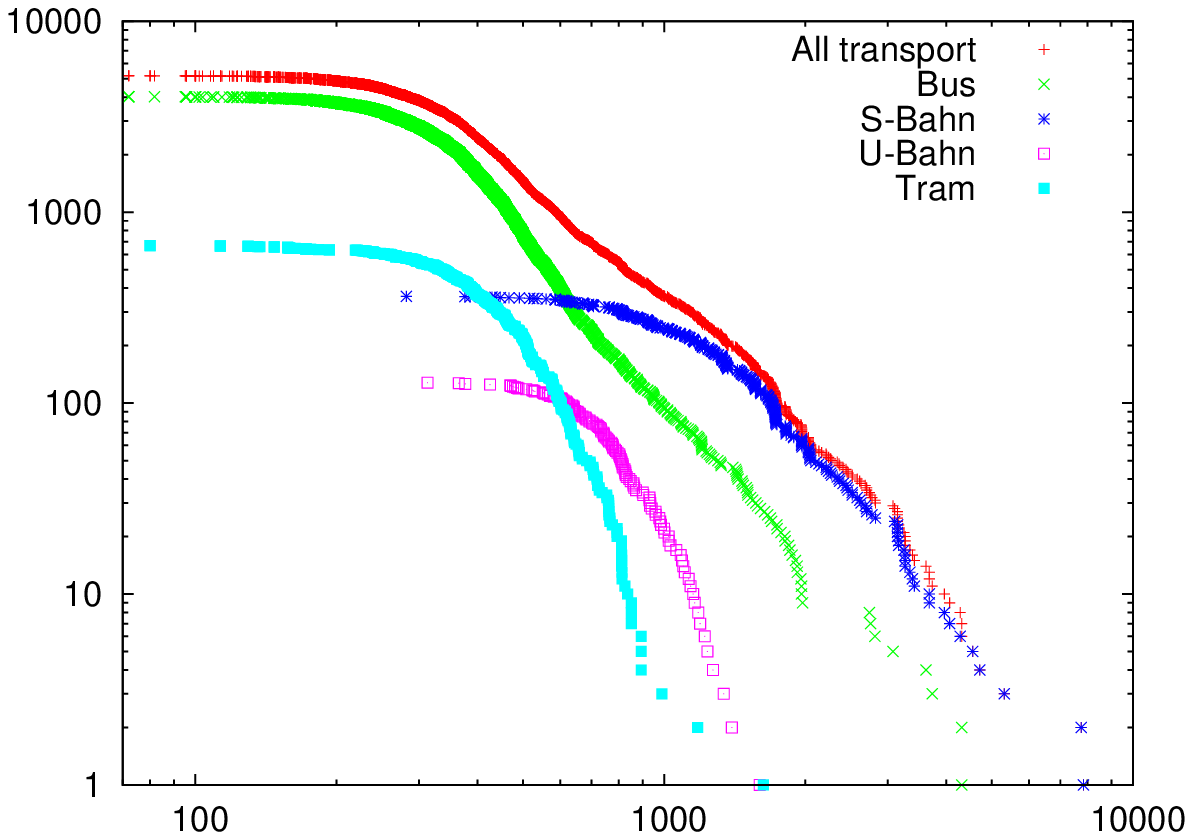}
\includegraphics[width=57mm]{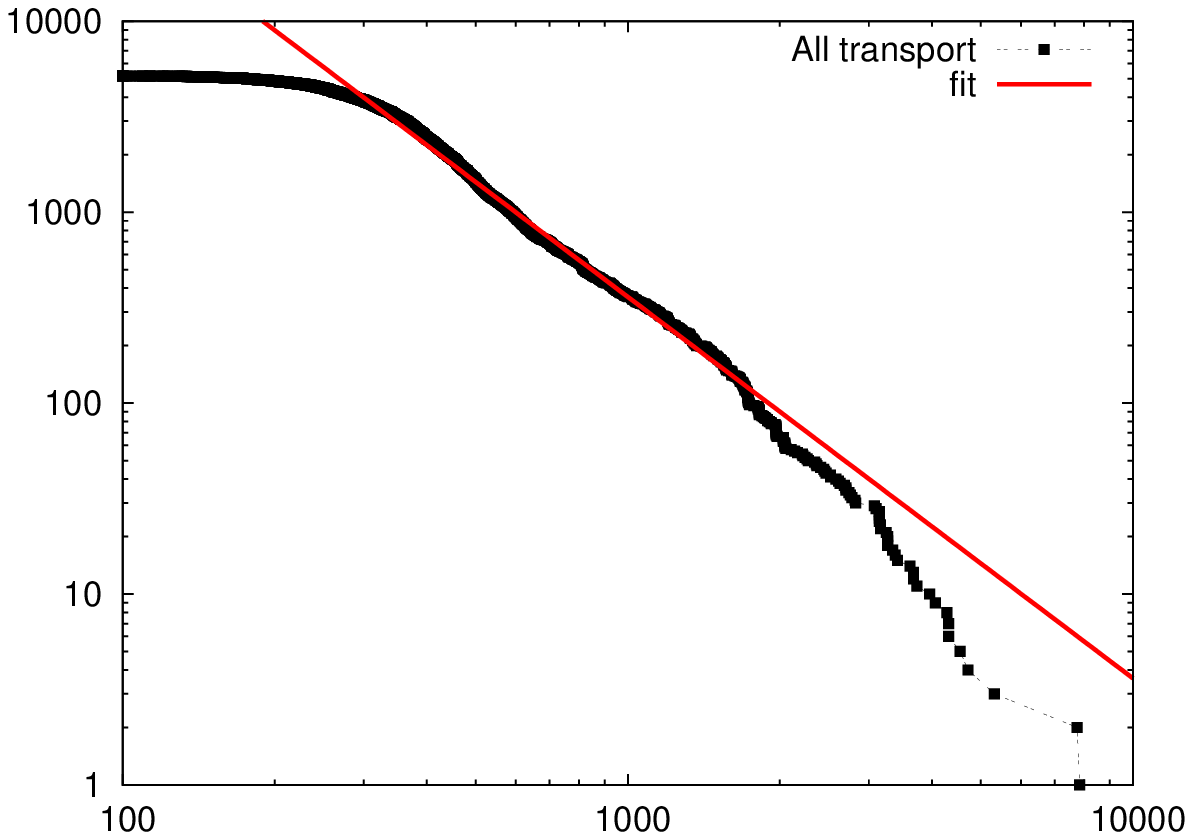}

\vspace*{1ex} {\bf a. \hspace{9em} b.}
 \caption{Cumulative
distribution for Berlin PTN: separate distributions for each means of transportation
({\bf a}) and the sum of these distributions ({\bf b}). The fit of the log-log plot in ({\bf b}) 
corresponds to an exponent value of $\mu =2 \pm 0.05$.}
 \label{fig1}
\end{figure}

To further analyze the situation,
let us explore in addition the statistics of the inter-station distances.
In Fig. \ref{fig1} we plot the cumulative distribution $P(\ell)$ of the
distances $\ell$ between consecutive stations along the routes showing
individual curves for each mode of transport as well as an integral curve
including all modes (Fig. \ref{fig1}{\bf a}). As follows
from the linear regions of the corresponding log-log plots, the region
of power law behaviour does not extend much more than one order of magnitude
for each single mode of transport. In the combined curve, this power law
regime is expanded.  The slopes and thus the corresponding exponents are
very similar for the bus and S-bahn routes. As one can deduce from  the plot,
the contribution of tram and U-bahn to the combined curve is rather weak.
In Fig. \ref{fig1}{\bf b} we show the fit of the combined curve to the power
law
\begin{equation} \label{3}
P(\ell)\sim \ell^{-\mu}
\end{equation}
with an appropriate exponent value
{\bf $\mu=2 \pm 0.05$}. Our finding is that at least in part
the distribution $P(\ell)$ displays scale-free properties.
In the following, we want to connect these scale-free properties observed for
large scales with the behaviour of the distribution for smaller distances $\ell$.
To this end we project the distance vectors $\underline{\ell}$ to the $x$ and $y$
axes and analyze the probability density functions (PDF) $p_x(x)$, $p_y(y)$.
Under the hypothesis that the tails of the distributions are governed by
a power law we may expect that the distributions $p_x(x)$,
$p_y(y)$ may be described by a L\'evy $\alpha$-stable distribution.
To fix notation we here specify the characteristic function of the
$\alpha$-stable L\'evy distribution
\begin{equation} \label{3a}
\hat{f}(t;\alpha,\beta,\tau,c) = \exp \Big [ -c^\alpha |t|^\alpha \Big (1-\lambda \beta {\rm sign} (t) \tan \frac{\pi\alpha}{2} \Big ) +i\tau t ) \Big ],
\hspace{1em} (\alpha\neq 1),
\end{equation}
with shape, skewness, location, and scale parameters $\alpha$, $\beta$, $\tau$, and $c$.
The asymptotic behaviour of the L\'evy $\alpha$-stable distribution
function $F(x;\alpha,\beta,\tau,c)$ for $\alpha<2$ is then given by \cite{Meyers10}:
\begin{equation} \label{4}
P(x) \simeq c^\alpha d (1+\beta) x^{-\alpha},\hspace{1em} \mbox{with} \hspace{1em} d=\sin (\pi\alpha/2) \Gamma (\alpha)/\pi.
\end{equation}
This relates the exponents, $\alpha=\mu$. Let us emphasize that the above fit concerns
the entire distribution function, and therefore it takes into account the behaviour at
smaller values of $\ell$. This is complementary to fitting
$P (\ell)$, where only the asymptotic (complementary) behaviour is taken into account.

\begin{figure}
\centering
\includegraphics[width=55mm]{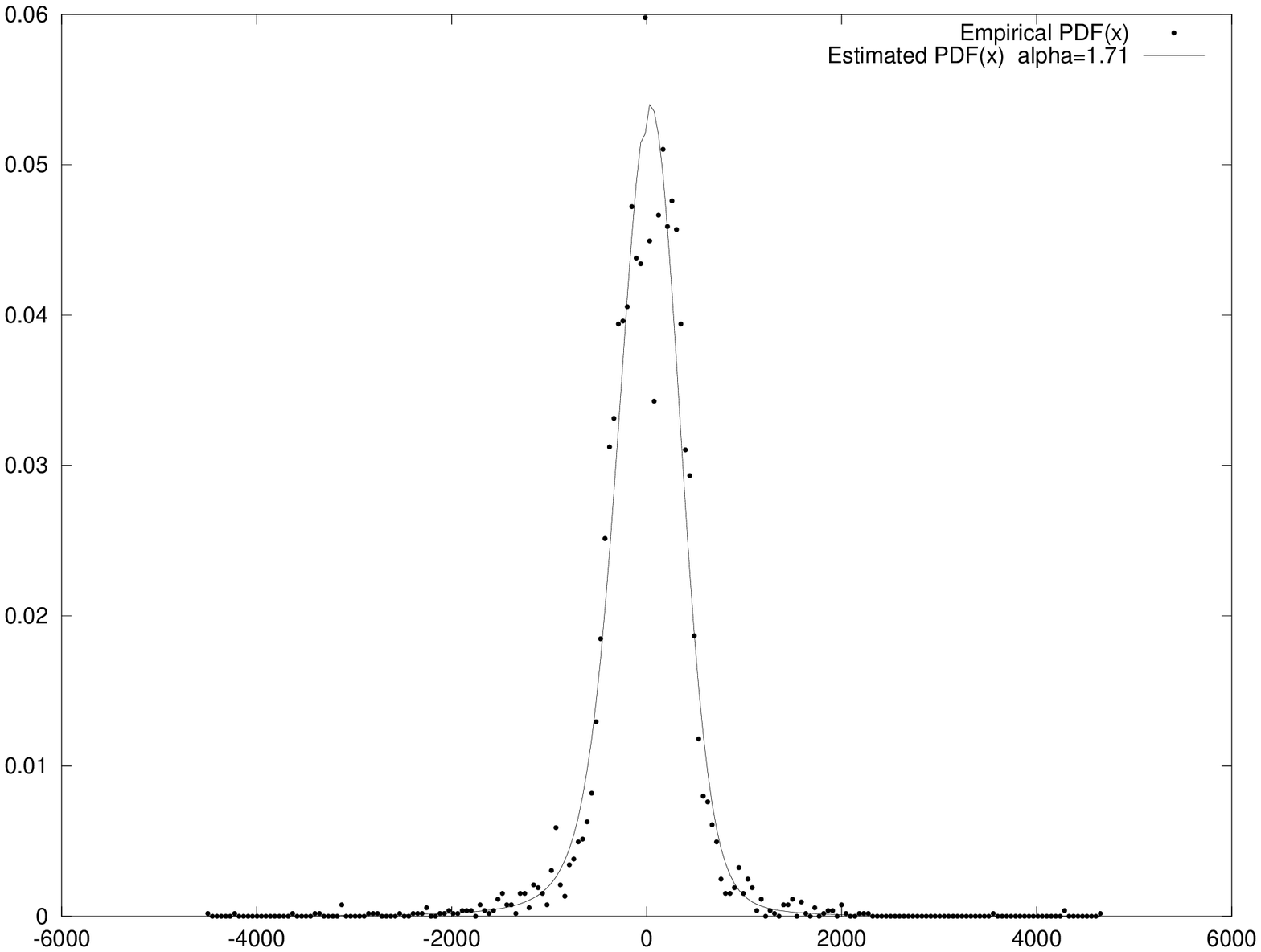}
\includegraphics[width=55mm]{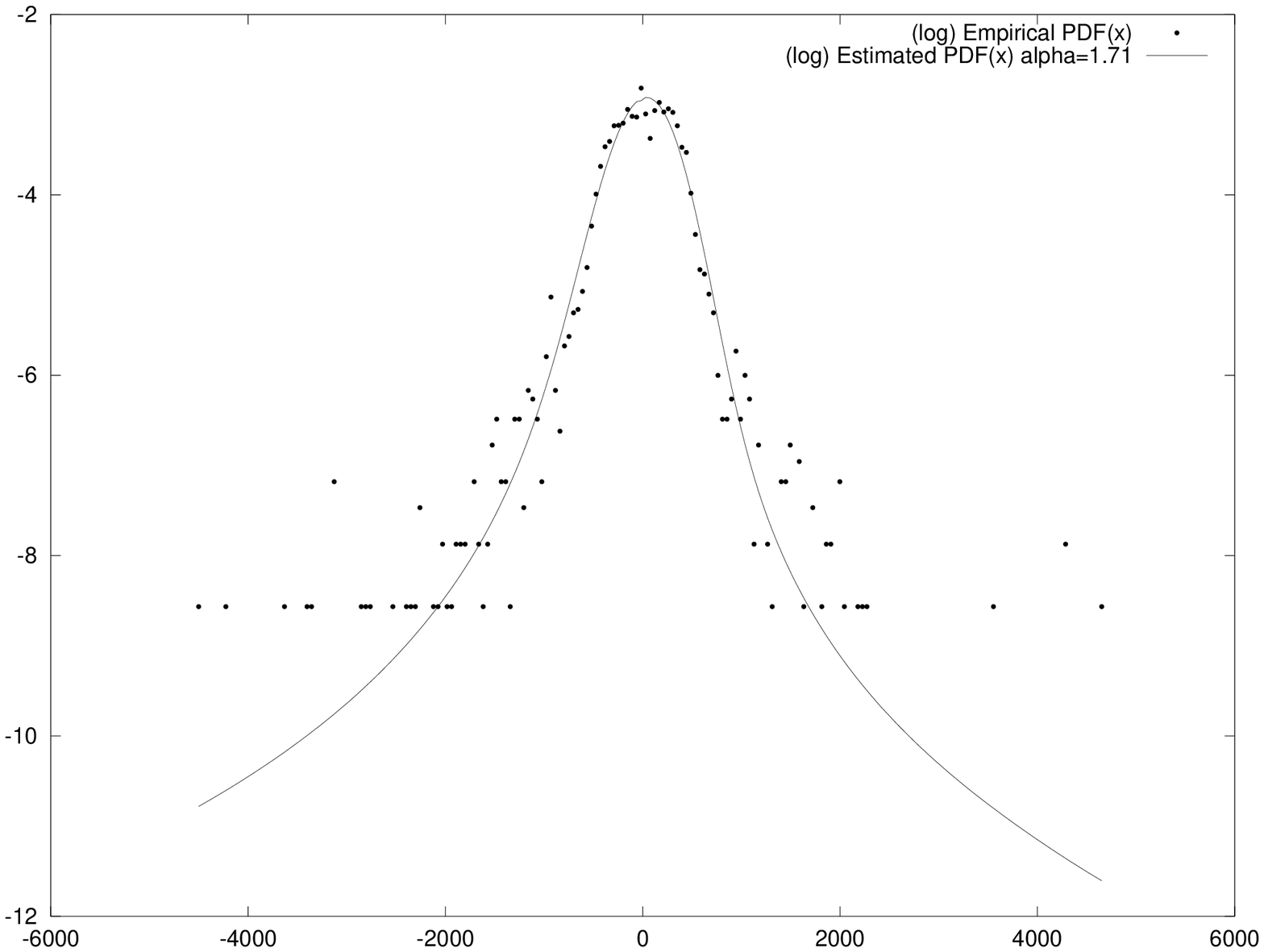}

\vspace*{1ex} {\bf a. \hspace{9em} b.} \vspace*{1ex}

\includegraphics[width=55mm]{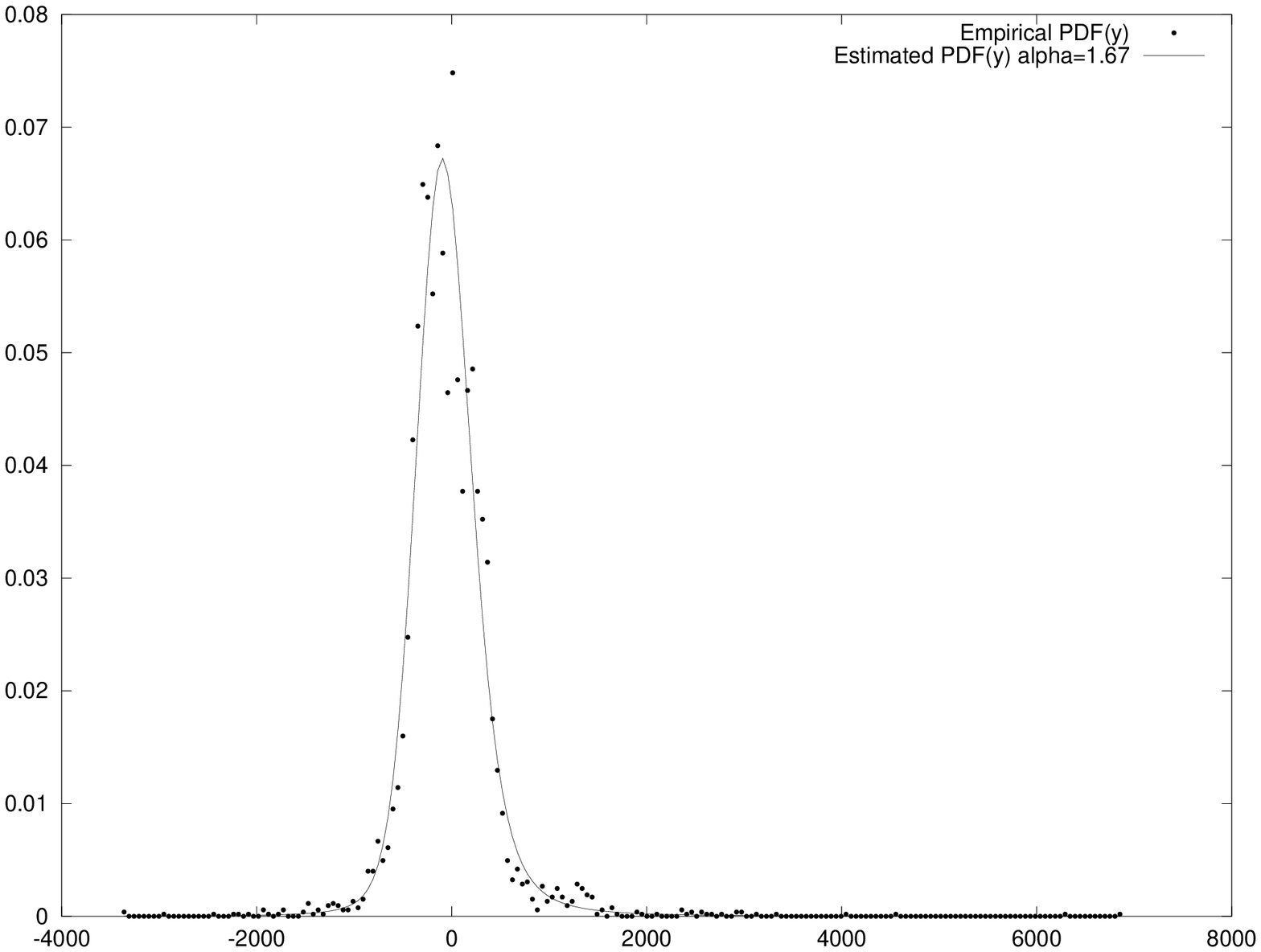}
\includegraphics[width=55mm]{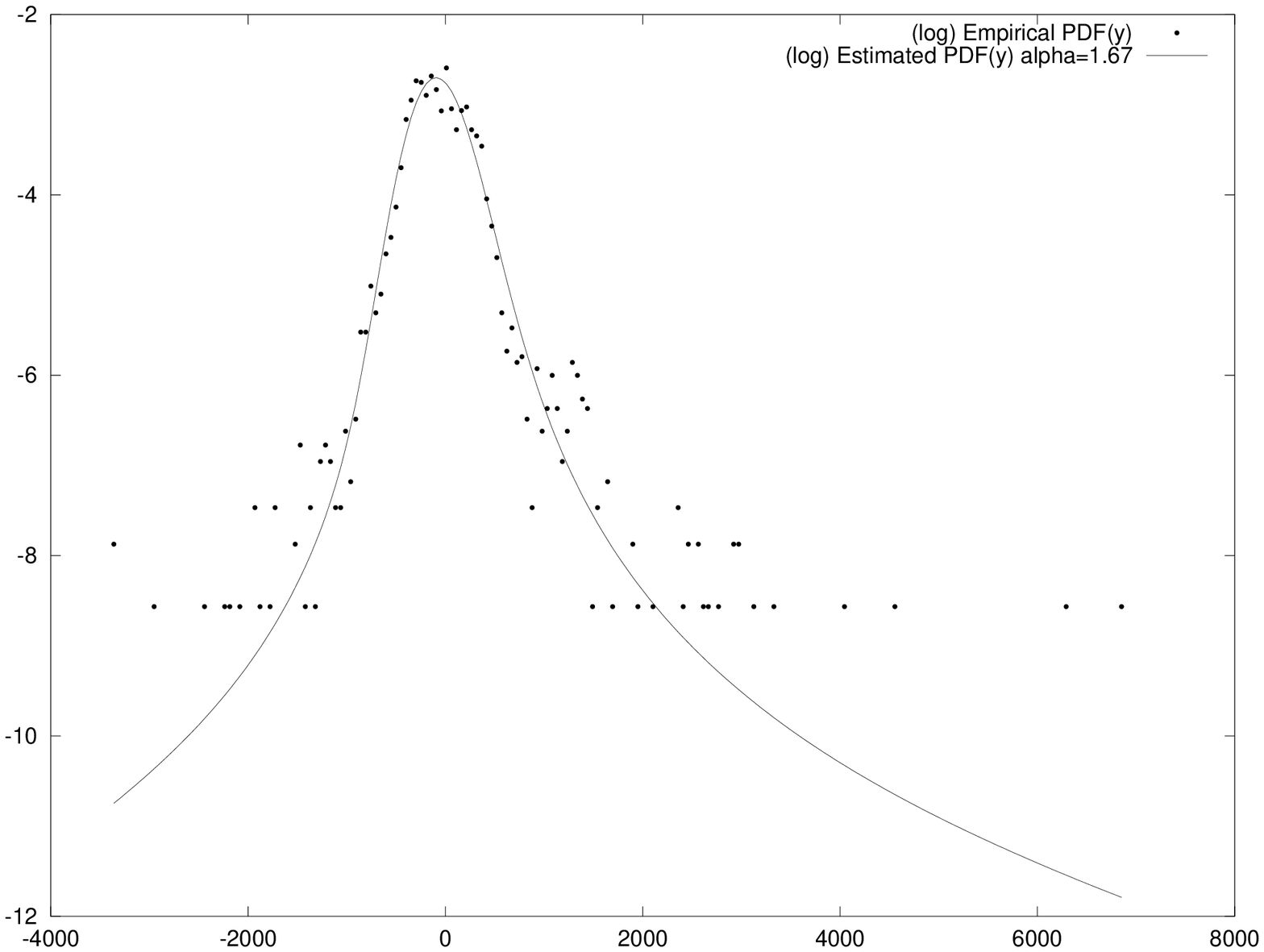}

\vspace*{1ex} {\bf c. \hspace{9em} d.}

\caption{Linear plots of the probability density functions (PDF) for
all means of transport projected to the $x$- and $y$- axes (left),
lin-log plots of the same (right) fitted to the $\alpha$-stable
distribution with $\alpha=1.71$ for the $x$-plots and $\alpha=1.67$
for the $y$-plots.}
 \label{fig2}
\end{figure}

In Fig. \ref{fig2} we fit the empirically determined PDFs $p_x(x)$,
$p_y(y)$ by an $\alpha$-stable distribution \cite{Veillette12}. The
best fits are obtained for values $\alpha=1.71$ for the
$x$-projection and $\alpha=1.67$ for the $y$-projection. These
values are to be compared with the value of $\mu$ derived above.
Recall that for a perfect $\alpha$-stable distribution the relation
$\alpha=\mu$ should hold. The latter holds in our case only
approximately. While one may not expect  high precision from the
given data, the obtained values of $\alpha$ and $\mu$ support the
conjecture that the inter-station distances are governed by a
non-trivial $\alpha$-stable distribution.

\begin{figure}
\centering
\includegraphics[width=55mm]{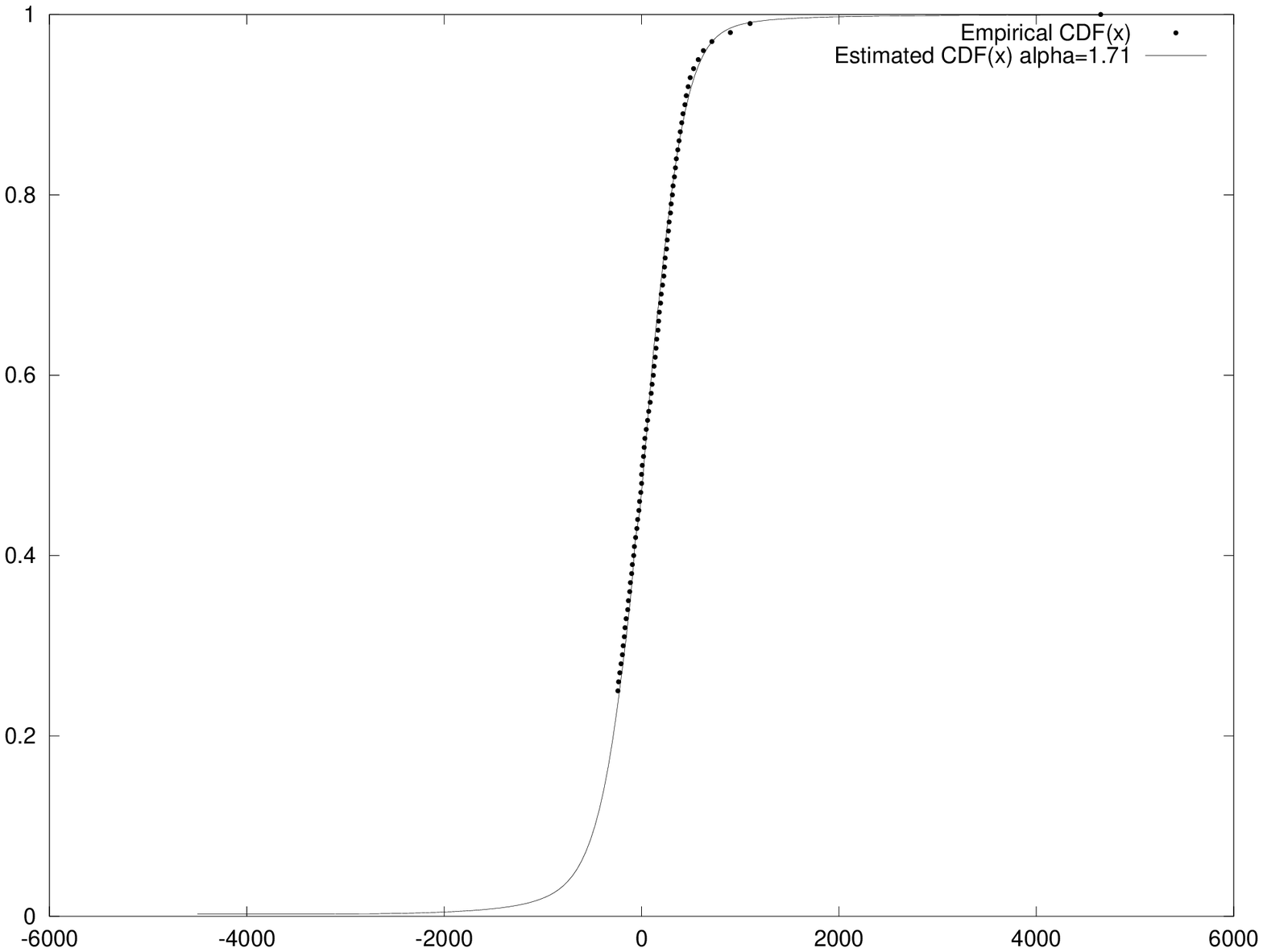}
\includegraphics[width=55mm]{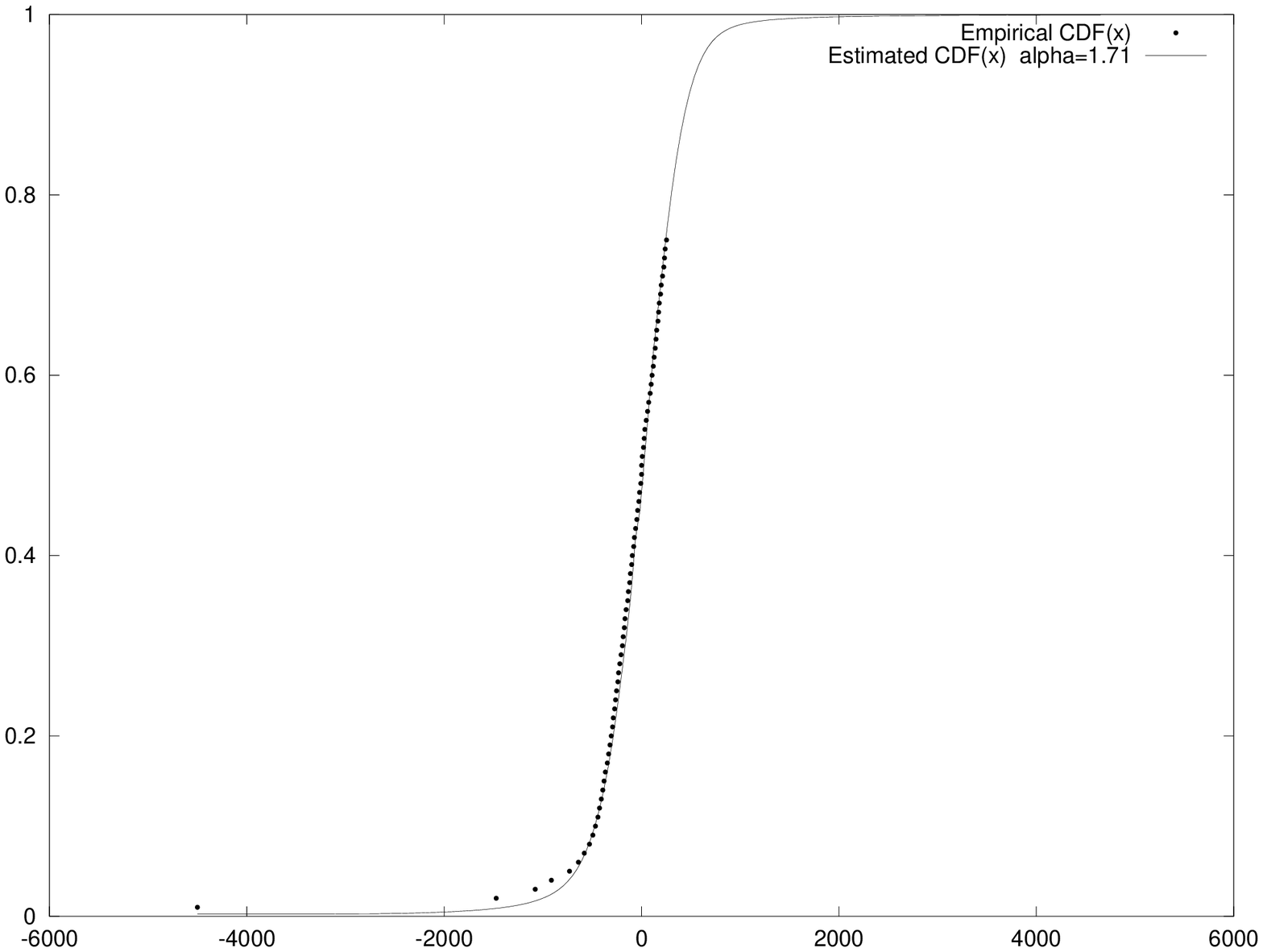}

\vspace*{1ex} {\bf a. \hspace{9em} b.} \vspace*{1ex}

\includegraphics[width=55mm]{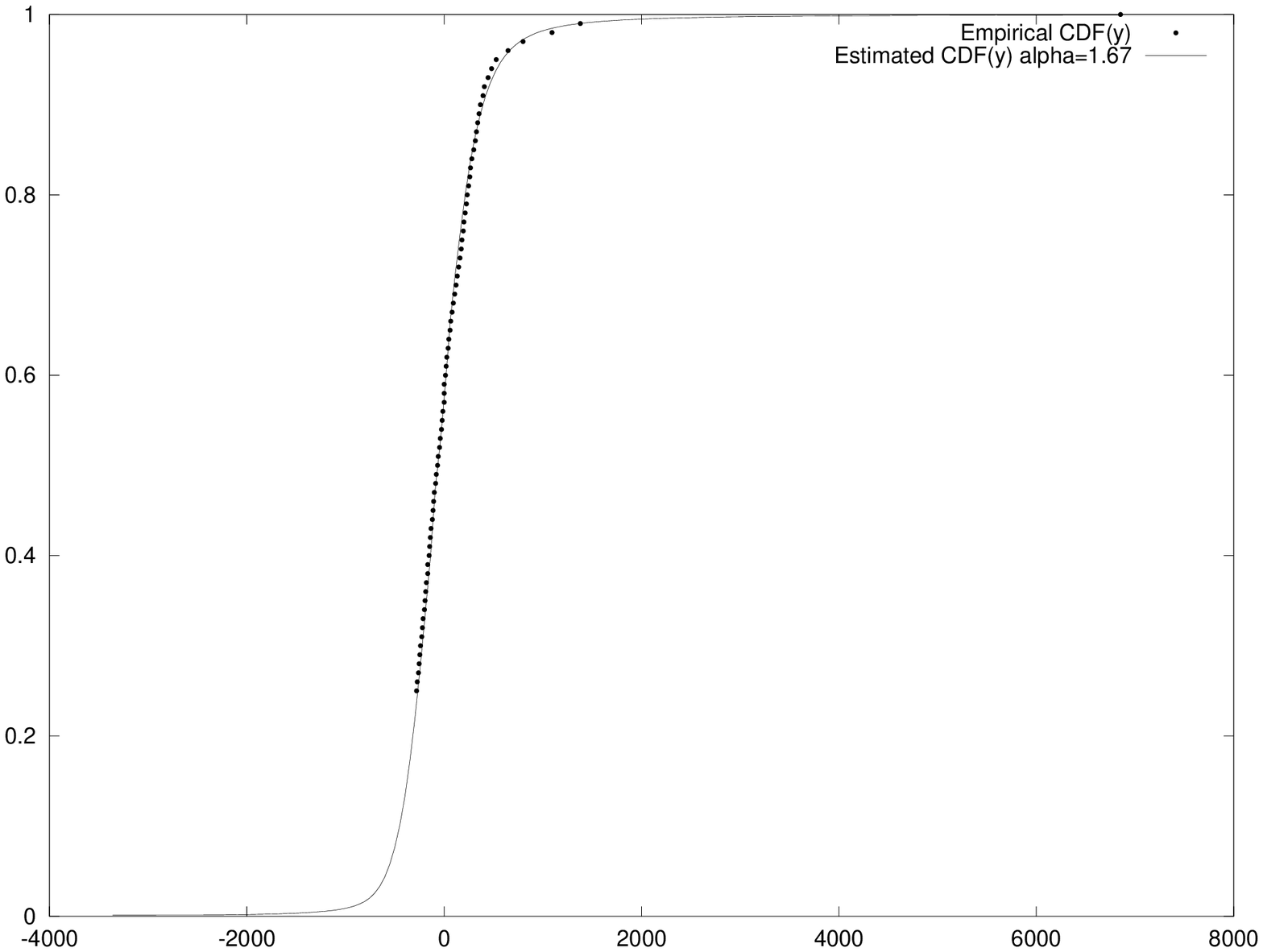}
\includegraphics[width=55mm]{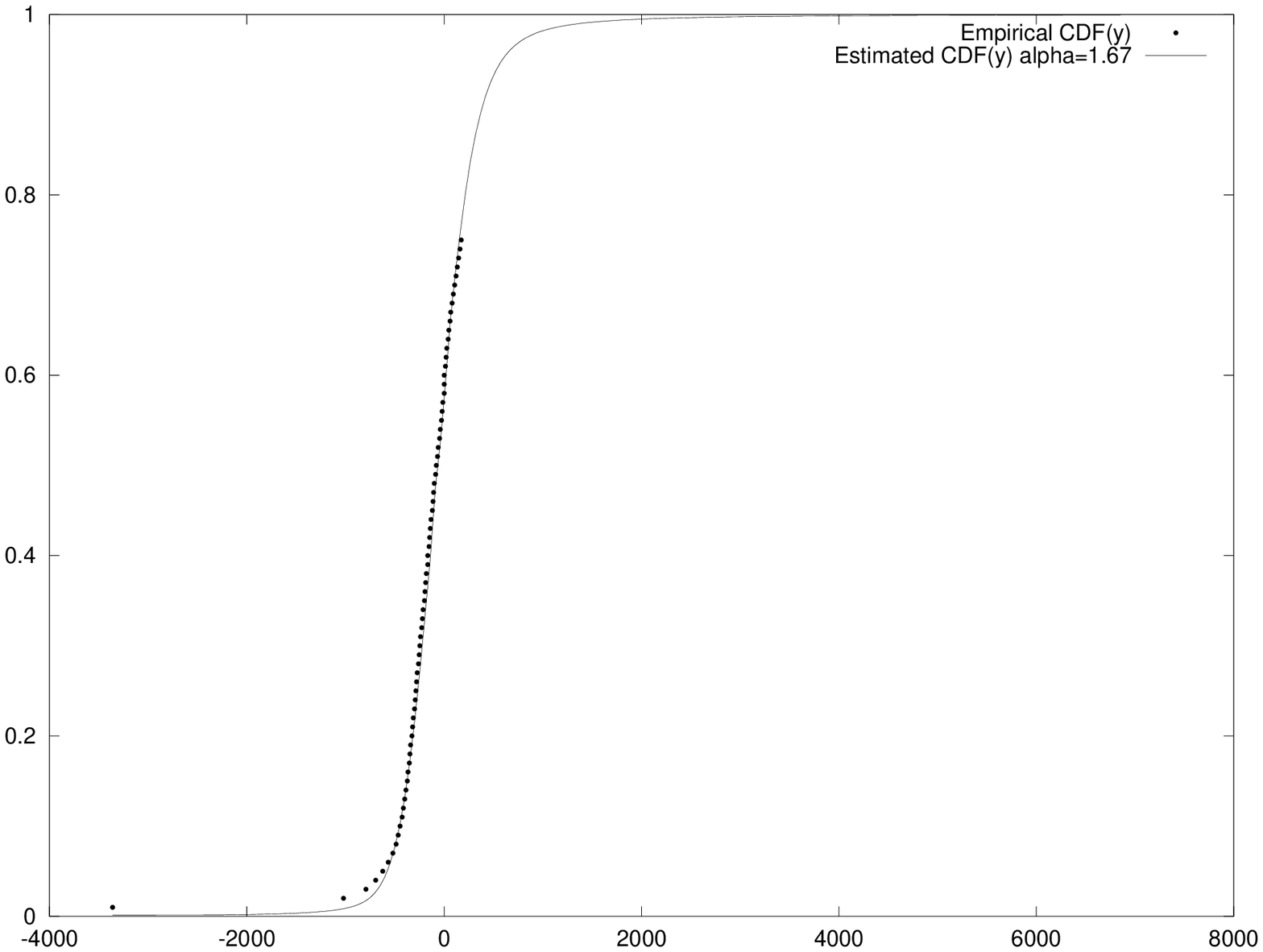}

\vspace*{1ex} {\bf c. \hspace{9em} d.} \vspace*{1ex}

\caption{Percentile plots of the cumulative $\alpha$-stable probability density distribution 
function (CDF) in comparison to the data. Distributions are shown in linear scale for the 
step-length projected to the $x$-axis (Figs 3a,b) with exponent $\alpha=1.71$ and to the
$y$-axis (Figs 3c,d) with  exponent $\alpha=1.67$. Figs. 3(a,c) display the behaviour of the 
25\%-100\% percentile while Figs. 3(b,d) display the 0\%-75\% percentile.
}
 \label{fig3}
\end{figure}

In further support of the fit to a L\'evy-stable distribution as described above, we
show in Fig. \ref{fig3} the so-called percentile plots for the
cumulative probability density functions (CDF) defined
as:
\begin{equation} \label{5}
P(x)=\int_{-\infty}^x p(x') dx'.
\end{equation}

In the light of these results, let us now re-examine the data on the end-to-end mean square inter-station
distance behaviour, cf. Eq. (\ref{2}). As mentioned above,
the power law scaling with the exponent $\nu$ has originally been interpreted in terms
of SAW behaviour in 2d. Note as well, that the
most reliable fit was achieved for the bus subnetwork which also provides the largest data set
and resulted in an exponent $\nu=0.82$.
Let us re-examine this value assuming as a working hypothesis a L\'evy flight
like behaviour for the inter-station distances.  Analytical and numerical results available
for self-avoiding L\'evy flights show that their scaling exponents differ from the usual
SAW exponents \cite{Grassberger85}. In particular, for the  node-avoiding L\'evy flight the
exponent $\nu$ is known in the form of  an $\varepsilon=(2\mu-d)$-expansion
\cite{Fisher72,Halley85}:
\begin{equation}\label{6}
\nu=\frac{1}{\mu} \Big ( 1 + \frac{1}{4\mu} \varepsilon  +
(19-\frac{5}{4}\mu^2)\frac{\varepsilon^2}{64\mu^2}+\cdots \Big ).
\end{equation}
Flory-type arguments on the other hand lead to \cite{Grassberger85}:
\begin{equation}\label{7}
\nu = \left \{ \begin{array}{ll}
 3/(\mu + d), & \quad d<d_c=2\mu \, , \\
 1/\mu, & \quad d>d_c .
\end{array}
\right .
\end{equation}
Substituting  the values $\mu=1.71$, $\mu=1.67$ into Eq. (\ref{6}) we get $\nu=0.80$, $\nu=0.81$,
whereas following Eq. (\ref{7}) we arrive at  $\nu=0.81$, $\nu=0.82$, both
compatible with the observed behaviour of the bus-subnetwork.

\section{Conclusions and outlook}
\label{III}
The present study extends previous observations of fractality within
public transport routes
\cite{vonFerber,Thibault87,Frankhauser90,Benguigui91,Benguigui92,Kim03}
by showing that also the distribution of
inter-station distances along routes displays non-trivial power law behaviour.
This indicates that the routes may in part also
be described as L\'evy-flights. The latter property may result
from the fact that the routes are planned to adapt to fluctuating
demand densities throughout the served area but also possibly related to
optimisation properties \cite{Humphries10}.

One may object, that the distribution derived above mixes different means of
transport with different mean inter-station distances. However, rather observing
journeys of individual passengers one may acknowledge that they will in
general use two or more means of transport to complete their journeys.
The given combined distribution of inter-station distances allows them
to perform this journey such that it resembles a L\'evy flight. This may allow
for travel optimization.

Self-avoiding L\'evy flights, apart from observing the constraint of non
self-intersec\-tion, evolve randomly. The fact that PT routes at least
within  the present sample appear to display the same scaling
symmetry is quite unexpected. In particular, this behavior seems to
be at odds with the requirement of minimizing passengers traveling
time between origin and destination. The latter argument, however,
ignores the time passengers spend walking to the initial and from
the final stations. Including these, one understands the need for
the routes to cover larger areas by meandering through
neighbourhoods. Given the requirements for a PTN to cover a
metropolitan area with a limited number of routes while
simultaneously offering fast transport across the city one may
speculate that routes scaling like self-avoiding L\'evy flights
may present an optimal solution \cite{Holovatch11,vonFerber_un}.

\section*{Acknowledgments}
This work was supported in part by the  7th FP, IRSES project
N269139 ``Dynamics and Cooperative phenomena in complex physical and
biological environments".

\end{document}